\documentclass[epj]{webofc}
\usepackage[utf8]{inputenc}
\usepackage[varg]{txfonts}   
\usepackage{booktabs}
\usepackage{xcolor}
\definecolor{darkred}{rgb}{0.4,0.0,0.0}
\definecolor{darkgreen}{rgb}{0.0,0.4,0.0}
\definecolor{darkblue}{rgb}{0.0,0.0,0.4}
\usepackage[bookmarks,linktocpage,colorlinks,
    linkcolor = darkred,
    urlcolor  = darkblue,
    citecolor = darkgreen]{hyperref}
\usepackage{subfigure}
\usepackage{amsmath}
\usepackage{amssymb}
\usepackage{bm}
\wocname{EPJ Web of Conferences}
\woctitle{Lattice2017}
\begin{document}
%
\selectlanguage{english}
\title{%
More on heavy tetraquarks in lattice QCD at almost physical pion mass
}
\author{%
\firstname{Anthony} \lastname{Francis}\inst{1}\fnsep\thanks{Speaker, \email{afranc@yorku.ca} } \and
\firstname{Renwick J.} \lastname{Hudspith}\inst{1} \and
\firstname{Randy} \lastname{Lewis}\inst{1} \and
\firstname{Kim}  \lastname{Maltman}\inst{1,2,3}
}
\institute{%
Department of Physics \& Astronomy, York University, 
Toronto, ON M3J 1P3, Canada
\and
Department of Mathematics \& Statistics, York University, 
Toronto, ON M3J 1P3, Canada
\and
CSSM, University of Adelaide, Adelaide SA 5005, Australia
}
\abstract{%
  We report on our progress in studying exotic, heavy tetraquark states, $qq’ \bar Q\bar Q’$. 
Using publicly available dynamical $n_f =2+1$ Wilson-Clover gauge configurations, generated by the PACS-CS collaboration, with pion masses $\simeq$164, 299 and 415 MeV, we extend our previous analysis to heavy quark components containing heavier than physical bottom quarks $\bar Q\bar Q’=\bar b'\bar b’$ or $\bar Q\bar Q’=\bar b\bar b’$, charm and bottom quarks $\bar c\bar b$ and also only charm quarks $\bar c\bar c$. Throughout we employ NRQCD and relativistic heavy quarks for the heavier than bottom, bottom and charm quarks.
Using our previously established diquark-antidiquark and meson-meson operator basis we comment in particular on the dependence of the binding energy on the mass of the heavy quark component $\bar Q\bar Q$, with heavy quarks ranging from $m_Q=0.85…6.3\cdot m_b$.
In the heavy flavor non-degenerate case, $\bar Q\bar Q’$, and especially for the tetraquark channel $ud\bar c\bar b$, we extend our work to utilize a $3\times 3$ GEVP to study the ground and threshold states thereby enabling a clear identification of possible binding.
Finally, we present initial work on the $\bar Q\bar Q’=\bar c\bar c$ system where a much larger operator basis is available in comparison to flavor combinations with NRQCD quarks.
}
\maketitle

\section{Introduction}\label{intro}




The study of exotic states in QCD can provide insights into the mechanisms behind the binding of quarks into hadrons. The only systematically-improvable way to investigate these is through an ab-initio procedure such as lattice QCD or by experiment. A benefit of the former is that arbitrary quark masses can be used as input into simulations to probe important features of composite states of quarks, such as the constituent quark mass dependencies, which will help us to understand why some configurations of quarks are bound while others are not.


Prior to our studies  \cite{Francis:2016nmj,Hudspith:2017bbh,Francis:2016hui} the question of whether bound heavy tetraquark states exist in nature had been tackled in lattice QCD mainly with heavy quarks in the static approximation and/or with very heavy light sea-quarks
\cite{Richards:1990xf,Mihaly:1996ue,Green:1998nt,Stewart:1998hk,Michael:1999nq,Pennanen:1999xi,Cook:2002am,Detmold:2007wk,Bali:2011gq,Brown:2012tm,Wagner:2011ev,Bicudo:2012qt,Ikeda:2013vwa,Guerrieri:2014nxa,Bicudo:2015vta,Bicudo:2016jwl}. 
In \cite{Francis:2016hui} we proposed an intuitive and phenomenologically-motivated binding mechanism for a $qq' \bar Q \bar Q'$ tetraquark with a heavy antidiquark $\bar Q \bar Q'$ component and a light $qq'$ part being in a good diquark configuation \cite{Jaffe:2004ph}. We studied the $qq'\bar Q\bar Q'$ = $ud\bar b\bar b$ and $\ell s\bar b\bar b$ channels, with nearly-physical dynamical light up and down ($l = u,d$) and strange quarks. The heavy bottom quarks were handled using lattice NRQCD. The binding energies obtained from a chiral extrapolation to the 
physical pion mass were $\Delta E_{ud\bar b \bar b}= 189(10)(3)$ MeV and $\Delta E_{ls\bar b \bar b}= 98(7)(3)$ MeV.

Since this prediction and the results from the static approximation\footnote{Both of which continue to be refined, see e.g. \cite{Peters:2016isf,Bicudo:2016ooe,Bicudo:2017szl}.}, compounded by the discovery of the doubly charmed $\Xi_{cc}$ baryon at LHCb \cite{Aaij:2017ueg}, there has been a renewed interest to explore this type of tetraquark configuration (see e.g. \cite{Karliner:2017qjm,Czarnecki:2017vco,Mehen:2017nrh}) and its quark mass dependence.

In this proceedings we report on progress of our own work to further understand the binding of heavy tetraquarks. We study the heavy quark mass dependence of the binding $\Delta E_{qq'\bar b \bar b'}(m_b,m_{b'})$ for unphysical bottom quark masses and also present first results for channels with the flavor configurations $ud\bar Q\bar Q' = ud \bar c\bar b$ and $ud \bar Q\bar Q' = ud\bar c\bar c$. The latter has also been investigated at $m_\pi=391$ MeV in \cite{Cheung:2017tnt} since this conference.

\section{Phenomenological considerations}
\label{sec:pheno}

To motivate the interpolating operators for our lattice calculation, observe that
in the limit of infinitely heavy quarks $m_{Q}\rightarrow\infty$ the attractive nature of the color Coulomb potential guarantees a bound ground state of a $qq' \bar Q \bar Q'$-type tetraquark \cite{Heller:1985cb}. Whether a binding is realized away from this limit, as in nature, for charm and bottom quarks, is subject to non-perturbative effects and only lattice QCD calculations can give a rigorous answer to this question.

However, there are indications from the observed spectrum that there should be tetraquark bound states
of the  $qq^\prime \bar b \bar b$-type\footnote{See e.g. \cite{Agashe:2014kda} as well as \cite{Francis:2016hui,Francis:2016nmj} and references therein for a list of model calculations on this topic.}, e.g. the mass ratios $(B^*-B)/(\Xi_{bb}^*-\Xi_{bb})$ and $(B_s^*-B_s)/(\Omega_{bb}^*-\Omega_{bb})$ are close to unity; this is indicative of the $b$-quark mass being large enough for heavy quark symmetry to be effective, which entails that its spin decouples and a heavy antidiquark in a color 3 configuration behaves similarly to a single heavy quark.
If heavy quark symmetry is indeed a good symmetry for bottom quarks, one might expect the observed heavy baryon spectrum gives an idea of the possible binding energies for tetraquarks, since the single heavy quark in the baryon may be replaced with a heavy anti diquark without changing the hadron's qualitative features. 
Through the splittings of the spin 0 and spin 1 diquark component baryons with the same flavor content, and a comparison to the corresponding spin averages, the possible binding energies could then be gauged.
In particular, we have  $\Sigma_b - \Lambda_b \approx 194$~MeV and $\Xi_b^\prime-\Xi_b\approx 162$~MeV \cite{Brown:2014ena}, i.e. the masses lie $\sim 145$~MeV below and $\sim 48$~MeV above the corresponding spin average in the $qq^\prime=ud$ case, and
$\sim 106$~MeV below and $\sim 35$~MeV above, for $qq^\prime=us$. In the  so-called "good diquark" spin 0 configuration \cite{Jaffe:2004ph} there is therefore an opportunity for binding energies in the same ballpark.

This motivation of a binding mechanism entails a number of predictions that may be tested using lattice calculations:
\begin{itemize}
\item The heavier the quarks in the $\bar Q \bar Q'$ component of the tetraquark candidate $qq'\bar Q \bar Q'$, the deeper the binding. The heavy quark mass dependence should be $\Delta E\sim 1/m_{\tilde Q}$.
\item The effectiveness of heavy quark symmetry is governed by the reduced mass of the two heavy quarks $\bar Q \bar Q'$.
\item The good diquark effect leads to a stronger binding for lighter quarks in the $qq'$ component of the tetraquark candidate $qq'\bar Q \bar Q'$.
\item There will be a maximum mass combination in $qq'$ above which the tetraquark candidate becomes unbound, if no further binding mechanisms become effective.
\end{itemize}
In this conference proceedings we report on our progress studying the first three of these predictions.

\section{Lattice correlators and operators}

The generic form of a lattice QCD correlation function at rest is given by
\begin{align}
C_{\mathcal{O}_1 \mathcal{O}_2}(t) = \sum_{{\bf x}} \Big\langle \mathcal{O}_1({\bf x},t) \mathcal{O}_2({\bf 0},0)^\dagger   \Big\rangle
=\sum_n \langle 0 | \mathcal{O}_1 | n \rangle\langle n| \mathcal{O}_2 | 0 \rangle e^{-E_n t}~~,
\end{align}
with the interpolating operators $\mathcal{O}_i$ being chosen with the quantum numbers of the continuum state to be studied. Given the phenomenological picture of Sec.~(\ref{sec:pheno}), we choose two types of operator:
First, we define a diquark-antidiquark operator of the form
\begin{equation}\label{eq:di_antidi}
\begin{aligned}
D(x) = ( u^\alpha_a(x) )^T ( C\gamma_5 )^{\alpha\beta} q^{\beta}_b(x) \; 
\times\;\bar{b}^\kappa_a(x) ( C\gamma_i )^{\kappa\rho} ( \bar{b}^\rho_b(x) )^T\;.
\end{aligned}
\end{equation}
This operator is expected to overlap with our possible tetraquark candidate as it has the light diquark in the favorable configuration. The natural alternative is a system of interacting mesons, which we implement using the following dimeson operator:
\begin{equation}\label{eq:dimeson}
\begin{aligned}
M(x) = 
\, \bar{b}^\alpha_a(x) \gamma_5^{\alpha\beta} u^{\beta}_a(x) \  
\bar{b}_b^{\kappa}(x) \gamma_i^{\kappa\rho} d_b^{\rho}(x)
\, - \bar{b}^\alpha_a(x) \gamma_5^{\alpha\beta} d^{\beta}_a(x)\  
\bar{b}_b^{\kappa}(x) \gamma_i^{\kappa\rho} u_b^{\rho}(x)\;.
\end{aligned}
\end{equation}
Both of these operators have the desired $J^P=1^+$ quantum numbers \cite{Francis:2016hui,Francis:2016nmj}.

Using this basis of operators, the energy spectrum of the given tetraquark channel may be extracted by first solving for the eigenvalues of the $2\times 2$ GEVP \cite{Blossier:2009kd}
\begin{equation}
F(t) = 
\begin{pmatrix}
G_{DD}(t) & G_{DM}(t)  \\
G_{MD}(t) & G_{MM}(t) 
\end{pmatrix},\quad F(t)\nu =\lambda(t) F(t_0) \nu\;,
\label{eq:gmatrix}
\end{equation}
where
\begin{equation}
G_{\mathcal{O}_1 \mathcal{O}_2} = \frac{C_{\mathcal{O}_1 \mathcal{O}_2}(t)}{C_{PP}(t) C_{VV}(t)}\;
\end{equation}
with $C_{PP}(t)$ and $C_{VV}(t)$ denoting the correlation functions of the individual pseudoscalar ($D,B,D_s,B_s$) and vector mesons ($D^*,B^*,D_s^*,B_s^*$), respectively. 
From the solution to the GEVP, the binding energy may be computed via a single-exponential fit to the obtained lowest lying, ground state, eigenvalue
\begin{equation}
\lambda_0(t) =  A e^{-\Delta E( t - t_0 ) }\;.
\end{equation}
In the case of non-degenerate heavy quarks $\bar Q\neq\bar Q'$ in $qq'\bar Q\bar Q'$ this correlator matrix may be extended to a $3\times 3$ GEVP, as a second threshold through a different flavor combination in the dimeson sector opens up: 
\begin{equation}
F(t) = 
\begin{pmatrix}
G_{DD} & G_{DM}  \\
G_{MD} & G_{MM} 
\end{pmatrix}~~ 
\Rightarrow
F(t) = 
\begin{pmatrix}
G_{DD} & G_{DM_{12}}  & G_{DM_{21}}  \\
G_{M_{12}D} & G_{M_{12}M_{12}} & G_{M_{12}M_{21}} \\
G_{M_{21}D} & G_{M_{21}M_{12}} & G_{M_{21}M_{21}}  \\
\end{pmatrix}~~.
\label{eq:3matrix}
\end{equation}
In the $ud\bar c\bar b$ channel these two thresholds are the $DB^*$ and $BD^*$, which are $\sim 97$ MeV \cite{Agashe:2014kda} apart.

\section{Numerical setup}

We use dynamical $n_f=2+1$ Wilson-Clover \cite{Sheikholeslami:1985ij} gauge field configurations generated by the PACS-CS collaboration \cite{Aoki:2008sm}, with a partially-quenched valence strange quark tuned to obtain the physical $K$ mass at the physical pion mass.
In the valence sector we use Coulomb gauge-fixed wall sources \cite{Hudspith:2014oja}. 
We set sources at multiple time positions and compute propagators for light and strange 
quarks using a modified deflated SAP-solver \cite{Luscher:2005rx}. 
An overview of the ensembles can be found in Tab.~\ref{tab:lat_par}(left) and further details may be found in \cite{Francis:2016hui,Francis:2016nmj,Hudspith:2017bbh}. 

\begin{table}[t!]
\centering
\begin{tabular}{c|ccc}
\hline\hline
& \multicolumn{3}{c}{Ensembles} \\
Label & $E_H$ & $E_M$& $E_L$ \\ \hline
Extent & $\:32^3\times64\:$ & $\:32^3\times64\:$ & $\:32^3\times64\:$ \\
$a^{-1}\;\left[\text{GeV}\right]$& 2.194(10) & 2.194(10) & 2.194(10) \\
$m_\pi L$ & 6.1 & 4.4 & 2.4 \\
$m_\pi\;\left[\text{MeV}\right]$ & 415 & 299 & 164 \\
$M_{\Upsilon}\;\left[\text{GeV}\right]$ & 9.528(79) & 9.488(71) & 9.443(76)\\
$M_{J/\Psi}\;\left[\text{GeV}\right]$ & 3.0862(2) & 3.0847(2) & 3.0685(11)\\
\hline\hline
\end{tabular}
\qquad\quad
\begin{tabular}{cc}
\hline\hline
\multicolumn{2}{c}{Heavy masses}\\
$m_{\rm bare}$ & $m_{b'}/m_{b}$ \\\hline
1.6 & 0.846(7) \\
3.0 & 1.463(12) \\
4.0 & 1.928(17) \\
8.0 & 4.395(35) \\
10.0 & 6.287(48)\\
\hline\hline
\end{tabular}
\caption{{Left: Overview of our ensemble parameters, more details may be found in the text and \cite{Francis:2016hui,Francis:2016nmj,Hudspith:2017bbh}. Right: Tuned unphysical heavy quark masses over the bottom quark mass. The tuning was achieved via the dispersion relation of spin-averaged mass mesons. These calculations were done using $E_M$ and one source position. }}
\label{tab:lat_par}
\end{table}

For the charm quarks we use a relativistic (Tsukuba-type) heavy quark action with tuning parameters taken from \cite{Namekawa:2011wt}
\begin{align}
D_{x,y}& = \delta_{xy} - \kappa_f \Big[ (1-\gamma_t)U_{x,t}\delta_{x+\hat t,y} + (1+\gamma_t)U_{x,t}\delta_{x+\hat t,y}  \Big] \\\nonumber
&-\kappa_f \sum_i \Big[ (r_s-\nu_s\gamma_i)U_{x,t}\delta_{x+\hat i,y} + (r_s+\nu_s\gamma_i)U_{x,t}\delta_{x+\hat i,y} \Big] 
-\kappa_f  \Big[ c_E \sum_{i} F_{it}(x)\sigma_{it} + c_B  \sum_{i,j} F_{ij}(x)\sigma_{ij}  \Big]~~.
\end{align}
Meson masses using quark propagators computed with this action and tuning are seen to be within $\sim 1\%$ of the experimentally observed spectrum. The implementation of this quark action is based once more on the openly available DDHMC package \cite{Luscher:2005rx}. 


To calculate bottom quark propagators we use the NRQCD lattice action with the Hamiltonian \cite{Thacker:1990bm,Davies:1994mp}
\begin{align}
H = &-\frac{\Delta^{(2)}}{2M_0}-c_1\frac{(\Delta^{(2)})^2}{8M_0^3}
+ \frac{c_2}{U_0^4}\frac{ig}{8M_0^2}(\bm{\tilde\Delta\cdot\tilde{E}}-\bm{\tilde{E}\cdot\tilde\Delta}) 
- \frac{c_3}{U_0^4}\frac{g}{8M_0^2}\bm{\sigma\cdot}(\bm{\tilde\Delta\times\tilde{E}}-\bm{\tilde{E}\times\tilde\Delta})\\\nonumber
&-\frac{c_4}{U_0^4}\frac{g}{2M_0}\bm{\sigma\cdot\tilde{B}}
+ c_5\frac{a^2\Delta^{(4)}}{24M_0}
- c_6\frac{a(\Delta^{(2)})^2}{16nM_0^2}\;,
\label{eq:nrqcd}
\end{align}
with the tadpole-improvement coefficient $U_0$ set to the fourth root of the plaquette and tree-level values $c_i=1$. A tilde denotes tree-level improvement and the $c_5,c_6$ terms remove the remaining $\mathcal{O}(a)$ and $\mathcal{O}(a^2)$ errors. This setup is known to account for relativistic effects at the few percent level while capturing the relevant heavy-light quark physics \cite{Lewis:2008fu,Gray:2005ur,Brown:2014ena}. 

In addition to tuning $M_0$ in the NRQCD action to achieve physical bottom quarks, we also varied this parameter to investigate heavier and lighter, unphysical bottom quark masses $m_{b^\prime}$ on the medium ensemble $E_M$, see Tab.~\ref{tab:lat_par}(right). 
We compute the slope of the spin-averaged mass dispersion relation to determine the (un)physical $b$ quark masses $m_{b^\prime}$. When compared to the physical bottom quark mass $m_b$ these are $m_{b^\prime}/m_b \approx 6.29, 4.40, 1.93, 1.46, 0.85$.
Static propagators were calculated by keeping only the first term in the NRQCD Hamiltonian.

\section{Results}

\subsection{Heavy quark mass dependence of $\Delta E_{ud\bar b \bar b'}(m_b,m_{b'})$ and $\Delta E_{ls\bar b \bar b'}(m_b,m_{b'})$ }

\begin{figure}[t!] 
  \centering
  \includegraphics[width=0.49\textwidth,clip]{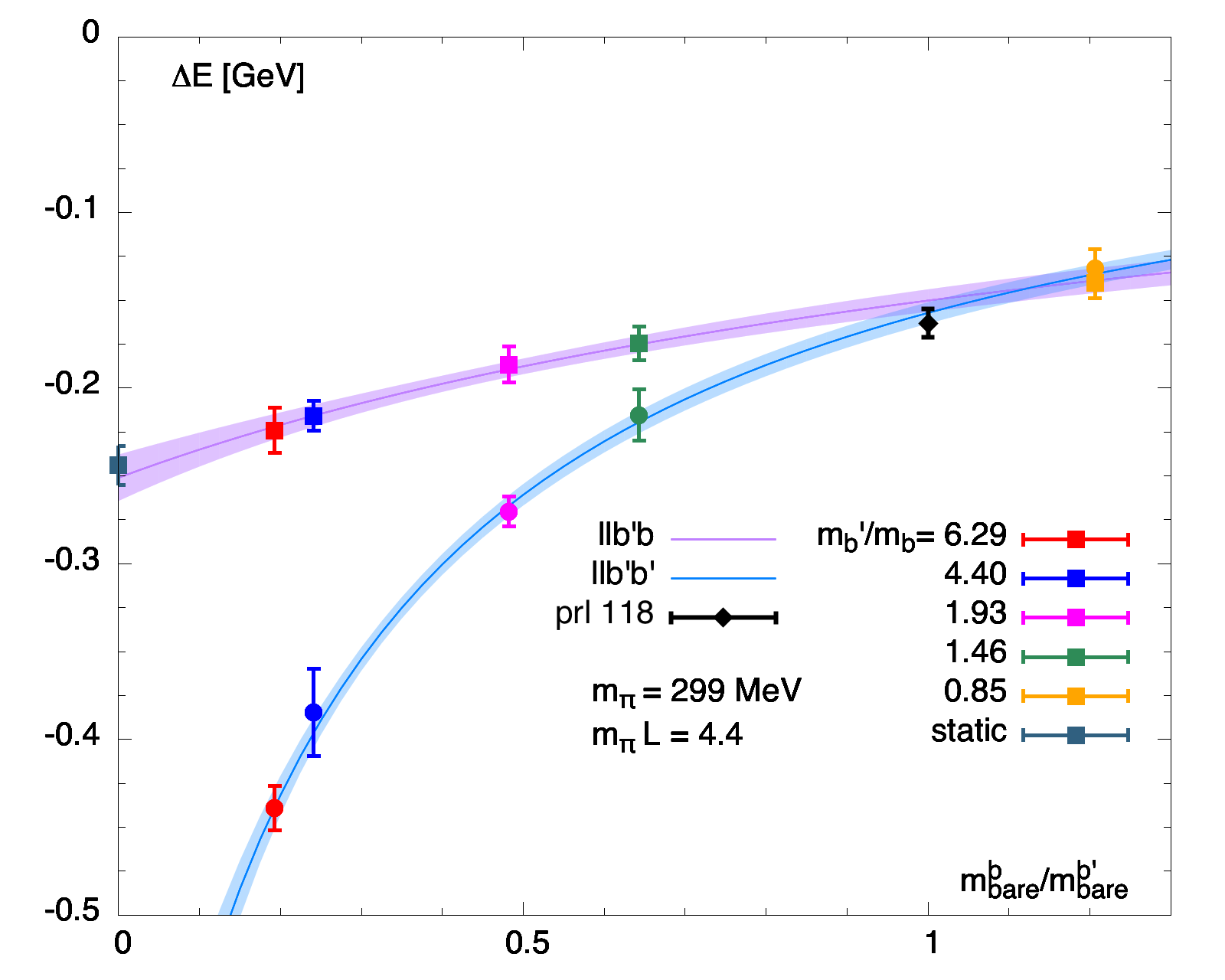}
  \includegraphics[width=0.49\textwidth,clip]{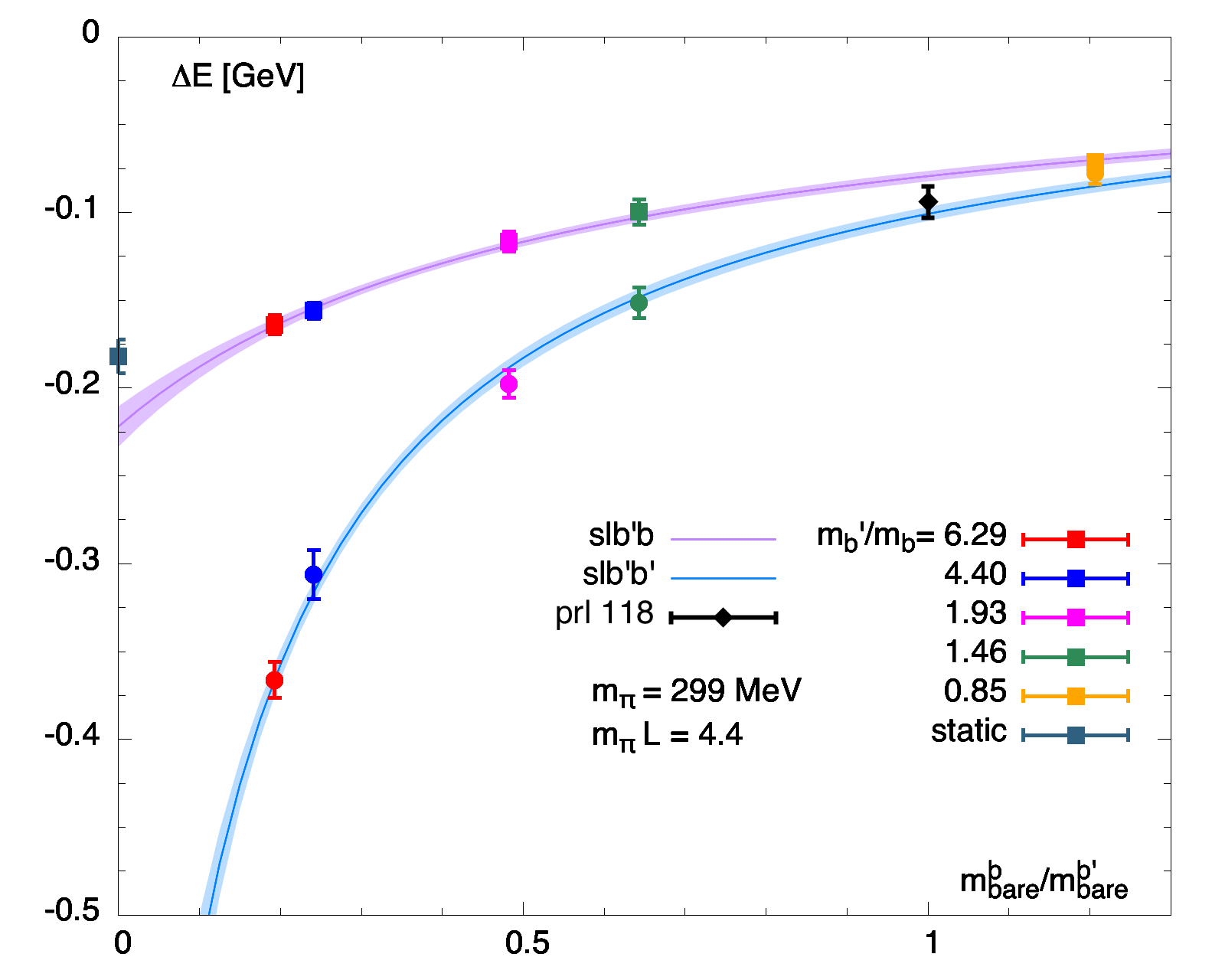}
  \caption{Results for the heavy quark mass dependence of the ${ud\bar b \bar b'}$ (left) and ${ls\bar b \bar b'}$ (right) tetraquark channels. In both figures the $m_b=m_{b'}$ and $m_b\neq m_{b'}$ mass dependences are shown. The extracted binding energies are fit to a $\sim 1/m_Q$ form. The results of \cite{Francis:2016hui} are given in black, they do not however enter the fit.}
  \label{fig:massdep}
\end{figure}

To test the heavy quark mass predictions of Sec.~(\ref{sec:pheno}) we proceed by calculating the binding energies $\Delta E_{ud\bar b \bar b'}(m_b,m_{b'})$ and $\Delta E_{ls\bar b \bar b'}(m_b,m_{b'})$ for the cases $m_b=m_{b'}$ and $m_b\neq m_{b'}$ for all available NRQCD heavy quark masses. As the extra, unphysical bottom quark masses are available only on $E_M$ a chiral, volume extrapolation is not possible. With $m_\pi L=4.4$ and based on the volume estimates of \cite{Francis:2016hui,Francis:2016nmj} the latter effects should be negligible. Taking the difference of $\Delta E_{\rm tetra}$ on $E_M$ and $E_{\rm phys}$ from \cite{Francis:2016hui} the former effect should be of the order $26~\rm MeV$ for ${ud\bar b \bar b}$ and $4~\rm MeV$ for ${ls\bar b \bar b}$. The results are shown in Fig.~(\ref{fig:massdep}), for the ${ud\bar b \bar b'}$ (left) and ${ls\bar b \bar b'}$ (right) cases. In both figures the $m_b=m_{b'}$ and $m_b\neq m_{b'}$ mass dependences are shown. In the $m_b\neq m_{b'}$ also the static propagators were used. The extracted binding energies are fit to a $\sim 1/m_Q$ form\footnote{Further details on fit ranges etc. will be given in an upcoming publication.}. 
Throughout, good agreement with the predicted $\sim 1/m_Q$ behavior is observed. This entails the binding mechanism (and heavy quark symmetry in particular) describes the observed behavior well in the mass region $m_Q=0.846m_b\rightarrow \infty$.

Note that the results of \cite{Francis:2016hui}, given in black, do not enter the fit. The predictions at the physical bottom quark mass are consistently postdicted by the $\sim 1/m_Q$ fit result.

\begin{figure}[t!] 
  \centering
  \includegraphics[width=0.49\textwidth,clip]{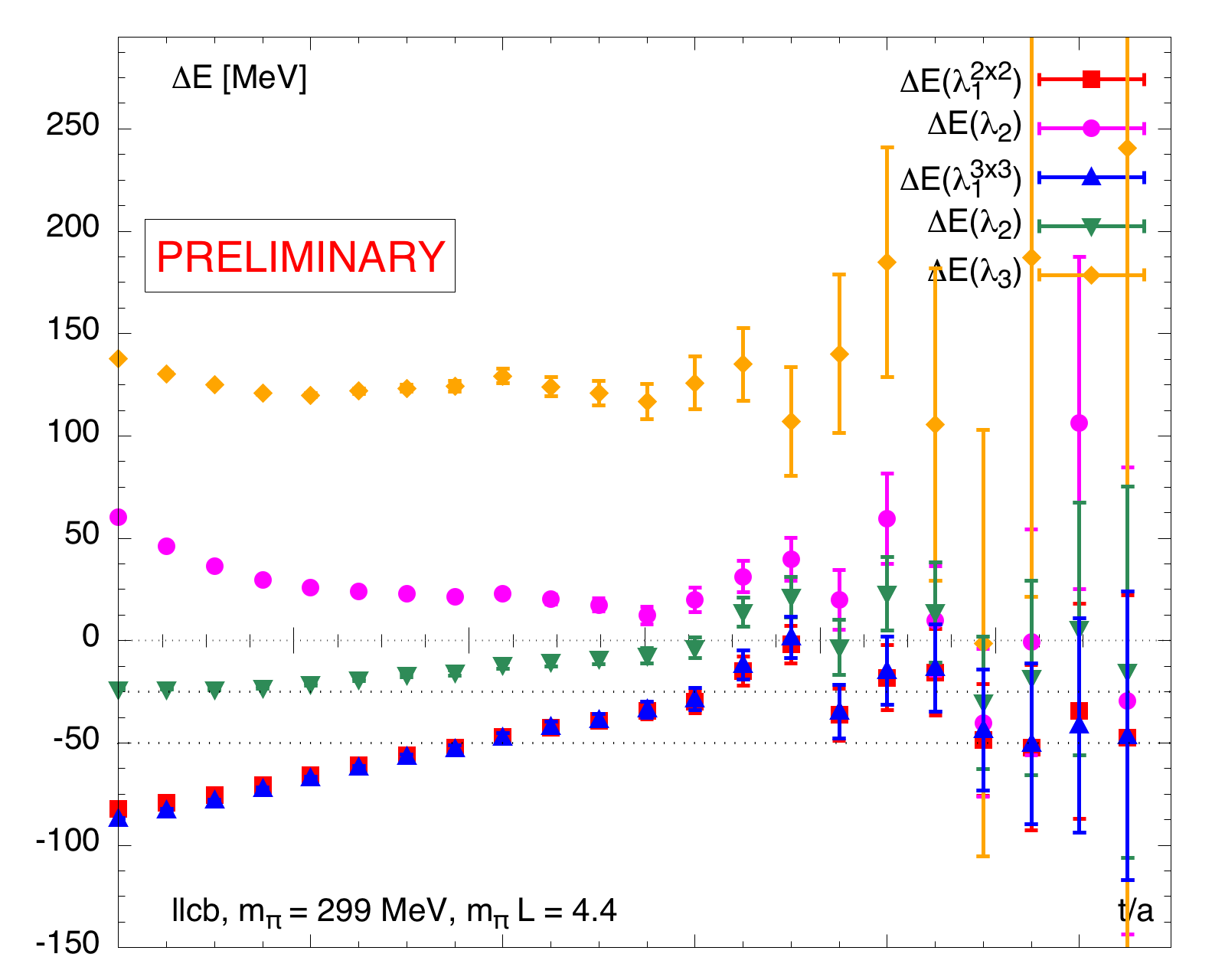}
  \includegraphics[width=0.49\textwidth,clip]{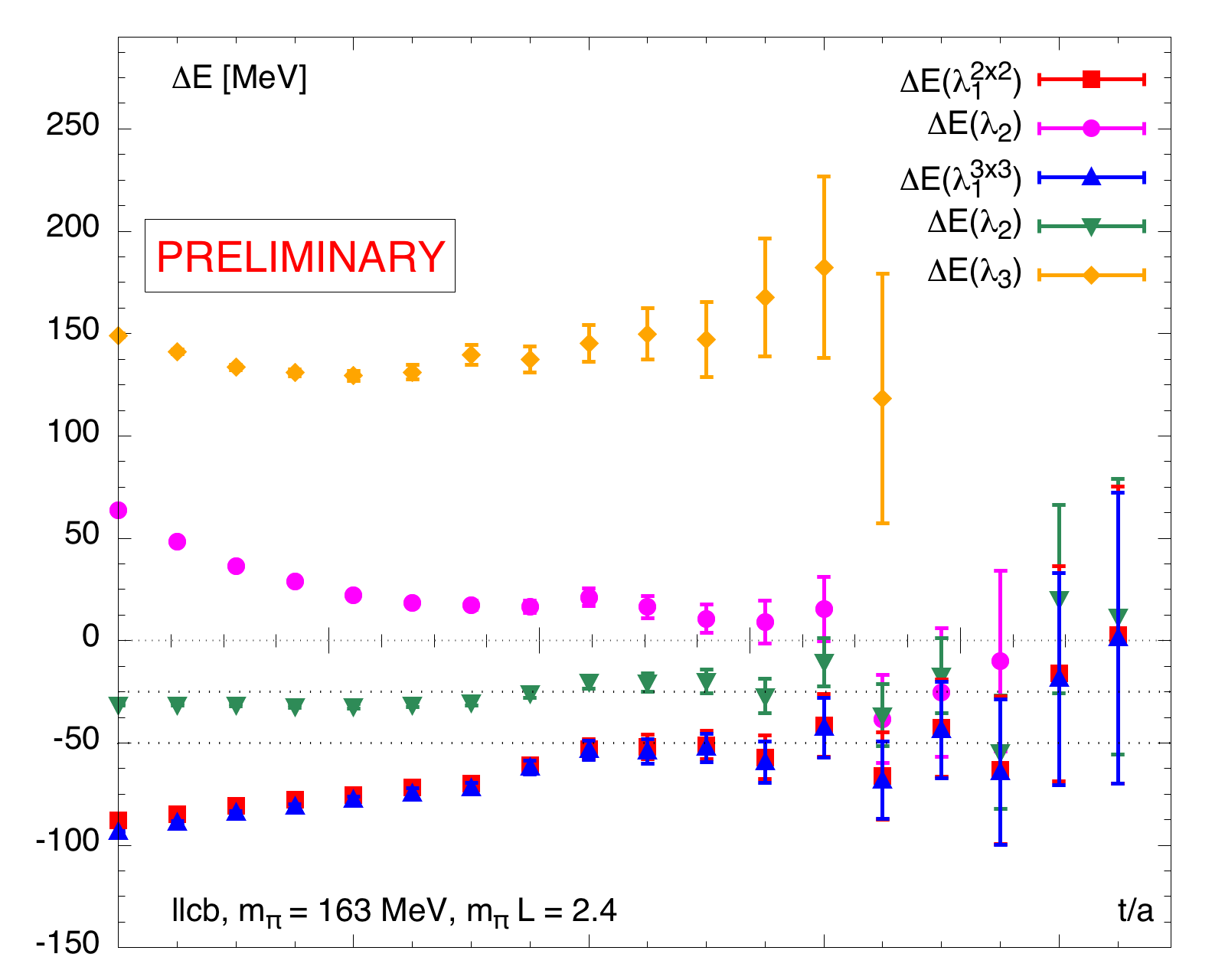}
  \caption{Preliminary results for the first charmed tetraquark candidate ${ud\bar c \bar b}$ on the medium $E_M$ (left) and light $E_L$ (right) ensembles. In both figures the results for the binding energies derived both from the $2\times 2$ and $3\times 3$ GEVP are given.  }
  \label{fig:udcb}
\end{figure}

\subsection{Evidence of binding in $ud\bar c \bar b$}

The results shown in Fig.~(\ref{fig:massdep}) give an indication that there may be further bound states as $m_{b^\prime} \rightarrow m_c$ and this motivates further investigation, although the case $m_{b^\prime} \approx m_c$ may be light enough for the binding and heavy quark symmetry arguments to break down.
With the prediction from heavy quark symmetry that the binding is dictated by the reduced mass of the $\bar Q\bar Q'$ component the first, most bound, likely tetraquark candidate in the charm quark mass regime is the flavor configuration ${ud\bar c \bar b}$. 
In Fig.~(\ref{fig:udcb}) we show preliminary results for this charmed-bottom tetraquark for the medium (left) and light (right) ensembles. This entails a light quark mass shift from $m_\pi=299$ MeV to $m_\pi=163$ MeV. In both figures the results for the binding energies derived both from the $2\times 2$ and $3\times 3$ GEVP are given.
As the binding mechanism predicts, and was observed in the ${ud\bar b \bar b}$ case, we expect the binding energy to increase for lighter quarks. Indeed, our results on the ensemble with almost physical quarks shows evidence of a binding at the $\mathcal{O}(50~\rm MeV)$-level with respect to the non-interacting two meson threshold. At $m_\pi=299$ MeV, however, the obtained signal is not as conclusive and permits an interpretation as being bound at the $\mathcal{O}(25~\rm MeV)$-level or indeed unbound. Both interpretations confirm the expectation of lighter diquark components binding more strongly. At the same time the results emphasize the importance of performing calculations at very light, preferably physical quark masses.

\subsection{First results for $ud\bar c \bar c$}

With the prediction of the binding energy set by the reduced mass of the heavy quark masses in $\bar Q\bar Q'$ of the tetraquark candidate, the above confirmation of this behavior in Fig.~(\ref{fig:massdep}) and the above
evidence of a bound ${ud\bar c \bar b}$ tetraquark Fig.~(\ref{fig:udcb}) further motivates the study of the $ud\bar c \bar c$ channel. It should be noted that since this conference results on this tetraquark candidate have been published \cite{Cheung:2017tnt}, in this study it was found to be unbound at quark masses corresponding to $m_\pi=391$ MeV. In light of the likely small or non-existent binding for ${ud\bar c \bar b}$ at $m_\pi=299$ MeV in our own calculation this highlights the necessity for almost physical quark masses for the study of these tetraquarks.

\begin{figure} 
  \centering
  \includegraphics[width=0.59\textwidth,clip]{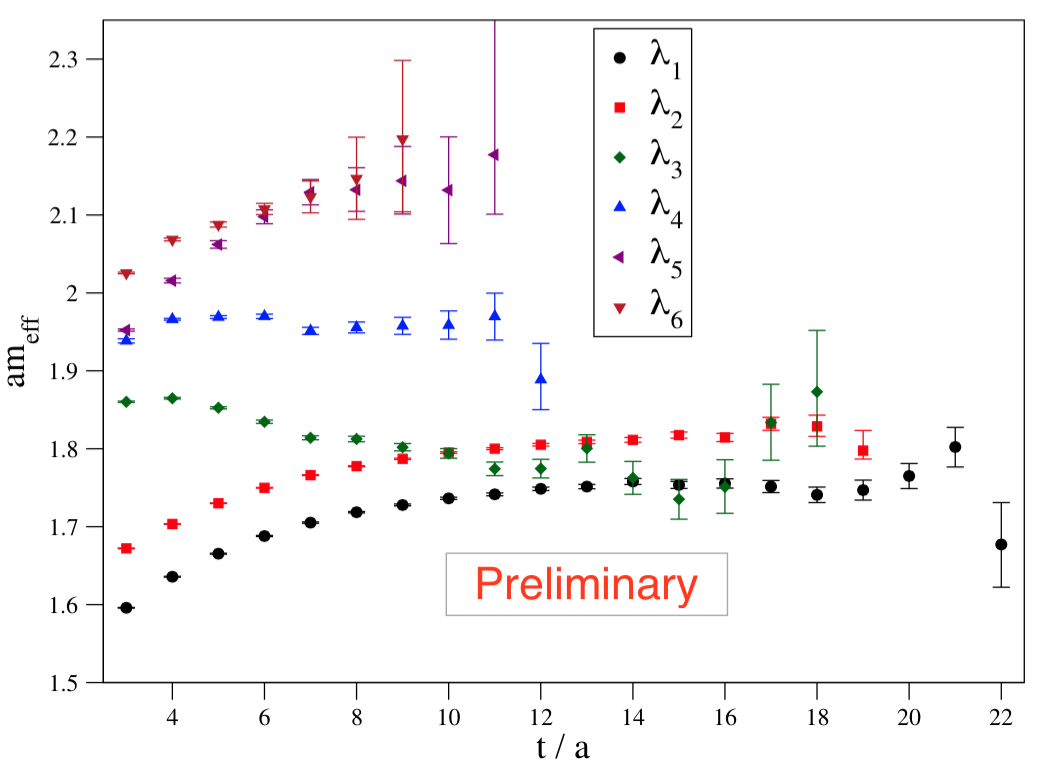}
  \caption{First preliminary results for a double charmed tetraquark candidate ${ud\bar c \bar c}$ on the medium $E_M$, $m_\pi=299$ MeV, ensemble. Without the need to use NRQCD much larger GEVPs may be set up, here we show results for a $6\times 6$ correlation matrix.}
  \label{fig:udcc}
\end{figure}

Numerically, with the absence of bottom quarks and therefore NRQCD propagators, the GEVP one may define is much larger than $3\times 3$, since many more operator combinations become available. In particular, we set up a $6\times 6$ GEVP of positive-positive and negative-negative parity operators in the diquark-diquark and dimeson-dimeson diagonal and mixing sectors. Preliminary results are shown in Fig.~(\ref{fig:udcc}) for the medium ensemble with $m_\pi=299$ MeV. 
Although strong conclusions cannot be drawn yet, especially lacking results which would allow us to perform the extrapolation to physical $m_\pi$, we do observe clear signals, and can successfully resolve the first three eigenvalues up to distances of $t/a=18$.

\section{Conclusions}
In this conference proceedings we report on progress in studying heavy tetraquarks of the ${qq'\bar Q\bar Q'}$-type. Following the predictions of the phenomenologically motivated binding mechanism through heavy quark symmetry and "good" diquarks, we can confirm numerically that the binding energy increases as $\sim 1/m_{\tilde Q}$ for the cases $m_Q=m_{Q'}$ and $m_Q\neq m_{Q'}$ between $m_{Q,Q'}\in[0.846\,m_b,\infty]$.
We see that the size of the heavy contribution to the total binding is dominated by the reduced mass of the two heavy quarks. As observed in our previous studies, the strongest binding contribution in the light sector is given for the $ud$ flavor configuration of the $qq'$ component of the tetraquark candidate. 
In addition we started to study the medium mass range by exchanging first one and then both of the bottom quarks by charm quarks, with the charm quarks handled using a relativistic heavy quark action. In the case of the ${ud\bar c \bar b}$ tetraquark channel, we find evidence of binding close to physical light quark masses. Although further investigation is required to pin down the binding energy, there is an indication of this state being bound at the $\Delta E_{ud\bar c \bar b} \sim 50~\rm MeV$-level. 
Exchanging also the second bottom quark with a charm quark, our initial results at $m_\pi=299$ MeV are not yet conclusive. However, should the observations made above hold, there is an indication for this tetraquark to also be bound.

For both the ${ud\bar c \bar b}$ and ${ud\bar c \bar c}$ channels more work is necessary in order to draw clear conclusions on their possible binding. Especially crucial here is the light quark mass dependence and physical or almost physical pion masses are mandatory. In addition, given the magnitudes of possible binding energies seen, finite volume effects play an important role in determining whether a genuine bound state or a scattering state is observed. Calculations at large lattice volumes are potentially necessary.

\section*{Acknowledgments}
The authors are supported by NSERC of Canada. Propagator inversions and gauge fixing were performed on Compute Canada's GPC machine at SciNet. Contractions were performed using our open-source contraction library \cite{contractions}.

%
%

%

%
%


\clearpage
\bibliography{lattice2017}

\begin{thebibliography}{42}

\bibitem{Francis:2016nmj}
A.~Francis, R.J. Hudspith, R.~Lewis, K.~Maltman, PoS \textbf{LATTICE2016}, 132
  (2016)

\bibitem{Hudspith:2017bbh}
R.J. Hudspith, A.~Francis, R.~Lewis, K.~Maltman, PoS \textbf{LATTICE2016}, 133
  (2017)

\bibitem{Francis:2016hui}
A.~Francis, R.J. Hudspith, R.~Lewis, K.~Maltman, Phys. Rev. Lett. \textbf{118},
  142001 (2017), \texttt{1607.05214}

\bibitem{Richards:1990xf}
D.G. Richards, D.K. Sinclair, D.W. Sivers, Phys. Rev. \textbf{D42}, 3191 (1990)

\bibitem{Mihaly:1996ue}
A.~Mihaly, H.R. Fiebig, H.~Markum, K.~Rabitsch, Phys. Rev. \textbf{D55}, 3077
  (1997)

\bibitem{Green:1998nt}
A.M. Green, P.~Pennanen, Phys. Rev. \textbf{C57}, 3384 (1998)

\bibitem{Stewart:1998hk}
C.~Stewart, R.~Koniuk, Phys. Rev. \textbf{D57}, 5581 (1998)

\bibitem{Michael:1999nq}
C.~Michael, P.~Pennanen, Phys. Rev. \textbf{D60}, 054012 (1999)

\bibitem{Pennanen:1999xi}
P.~Pennanen, C.~Michael, A.M. Green, Nucl. Phys. Proc. Suppl. \textbf{83}, 200
  (2000)

\bibitem{Cook:2002am}
M.S. Cook, H.R. Fiebig (2002)

\bibitem{Detmold:2007wk}
W.~Detmold, K.~Orginos, M.J. Savage, Phys. Rev. \textbf{D76}, 114503 (2007)

\bibitem{Bali:2011gq}
G.~Bali, M.~Hetzenegger, PoS \textbf{LATTICE2011}, 123 (2011)

\bibitem{Brown:2012tm}
Z.S. Brown, K.~Orginos, Phys. Rev. \textbf{D86}, 114506 (2012)

\bibitem{Wagner:2011ev}
M.~Wagner, Acta Phys. Polon. Supp. \textbf{4}, 747 (2011)

\bibitem{Bicudo:2012qt}
P.~Bicudo, M.~Wagner, Phys. Rev. \textbf{D87}, 114511 (2013)

\bibitem{Ikeda:2013vwa}
Y.~Ikeda et~al., Phys. Lett. \textbf{B729}, 85 (2014)

\bibitem{Guerrieri:2014nxa}
A.L. Guerrieri, et~al., PoS \textbf{LATTICE2014}, 106 (2015)

\bibitem{Bicudo:2015vta}
P.~Bicudo, K.~Cichy, A.~Peters, B.~Wagenbach, M.~Wagner, Phys. Rev.
  \textbf{D92}, 014507 (2015)

\bibitem{Bicudo:2016jwl}
P.~Bicudo, J.~Scheunert, M.~Wagner, PoS \textbf{LATTICE2016}, 103 (2016),
  \texttt{1609.00548}

\bibitem{Jaffe:2004ph}
R.L. Jaffe, Phys. Rept. \textbf{409}, 1 (2005)

\bibitem{Peters:2016isf}
A.~Peters, et~al., PoS \textbf{LATTICE2016}, 104 (2016), \texttt{1609.00181}

\bibitem{Bicudo:2016ooe}
P.~Bicudo, J.~Scheunert, M.~Wagner, Phys. Rev. \textbf{D95}, 034502 (2017),
  \texttt{1612.02758}

\bibitem{Bicudo:2017szl}
P.~Bicudo, et~al., Phys. Rev. \textbf{D96}, 054510 (2017), \texttt{1704.02383}

\bibitem{Aaij:2017ueg}
R.~Aaij et~al. (LHCb), Phys. Rev. Lett. \textbf{119}, 112001 (2017),
  \texttt{1707.01621}

\bibitem{Karliner:2017qjm}
M.~Karliner, J.L. Rosner (2017), \texttt{1707.07666}

\bibitem{Czarnecki:2017vco}
A.~Czarnecki, B.~Leng, M.B. Voloshin (2017), \texttt{1708.04594}

\bibitem{Mehen:2017nrh}
T.~Mehen (2017), \texttt{1708.05020}

\bibitem{Cheung:2017tnt}
G.K.C. Cheung, et~al. (Hadron Spectrum) (2017), \texttt{1709.01417}

\bibitem{Heller:1985cb}
L.~Heller, J.A. Tjon, Phys. Rev. \textbf{D32}, 755 (1985)

\bibitem{Agashe:2014kda}
K.A. Olive et~al. (Particle Data Group), Chin. Phys. \textbf{C38}, 090001
  (2014)

\bibitem{Brown:2014ena}
Z.S. Brown, W.~Detmold, S.~Meinel, K.~Orginos, Phys. Rev. \textbf{D90}, 094507
  (2014)

\bibitem{Blossier:2009kd}
B.~Blossier, et~al., JHEP \textbf{04}, 094 (2009), \texttt{0902.1265}

\bibitem{Sheikholeslami:1985ij}
B.~Sheikholeslami, R.~Wohlert, Nucl. Phys. \textbf{B259}, 572 (1985)

\bibitem{Aoki:2008sm}
S.~Aoki et~al., Phys. Rev. \textbf{D79}, 034503 (2009)

\bibitem{Hudspith:2014oja}
R.J. Hudspith, Comput. Phys. Commun. \textbf{187}, 115 (2015)

\bibitem{Luscher:2005rx}
M.~L{\"u}scher, Comput.Phys.Commun. \textbf{165}, 199 (2005)

\bibitem{Namekawa:2011wt}
Y.~Namekawa et~al. (PACS-CS), Phys. Rev. \textbf{D84}, 074505 (2011),
  \texttt{1104.4600}

\bibitem{Thacker:1990bm}
B.A. Thacker, G.P. Lepage, Phys. Rev. \textbf{D43}, 196 (1991)

\bibitem{Davies:1994mp}
C.T.H. Davies, et~al., Phys. Rev. \textbf{D50}, 6963 (1994)

\bibitem{Lewis:2008fu}
R.~Lewis, R.M. Woloshyn, Phys. Rev. \textbf{D79}, 014502 (2009)

\bibitem{Gray:2005ur}
A.~Gray, et~al., Phys. Rev. \textbf{D72}, 094507 (2005)

\bibitem{contractions}
\url{https://github.com/RJHudspith/Contractual\_Obligations}

\end{thebibliography}

\end{document}